\documentclass[aps,twocolumn]{revtex4}

\usepackage{amsmath}
\usepackage{amssymb}
\usepackage[dvips]{graphicx}

\begin{document}

\title{A Comment on Bonnor-Steadman  Closed Timelike Curves. }

\author{Val\'eria M. Rosa\footnote{e-mail: vmrosa@ufv.br}
 }

\affiliation{
Departamento de Matem\'atica, Universidade Federal de Vi\c{c}osa,
36570-000 Vi\c{c}osa, M.G., Brazil
}

\author{ Patricio  S. Letelier\footnote{e-mail: letelier@ime.unicamp.br}
}

\affiliation{
Departamento de Matem\'atica Aplicada-IMECC,
Universidade Estadual de Campinas,
13081-970 Campinas,  S.P., Brazil}
  
\begin{abstract}
 The existence and stability closed timelike curves in a Bonnor-Ward
 spacetime without torsion line singularities is shown by
 exhibiting particular examples. 
\end{abstract}
\maketitle
In a seminal paper Bonnor and Steadman \cite{bonnorst} study the existence 
of closed timelike curves (CTC) in the Bonnor-Ward metric \cite{bonnorwd} 
whose source are two spinning charged particles  
with masses $m_1$ and $m_2,$ and magnetic
 moments $\mu_1$ and $\mu_2$ placed on the $z$-axis at $z=\pm a,\,(a>0)$.
 The source is bounded and the spacetime is asymptotically flat. 
In principle,  this spacetime is physically realizable 
\cite{bonnorst}. Furthermore, in  this spacetime we have 
regions with stable closed timelike geodesics (CTGs) \cite{rosalet}. 

In the spacetime for the already studied cases of CTCs and CTGs,
besides the singularities associated the punctual masses that in the
particular examples studied are equal, we ought to have a torsion line
singularity (TLS)\cite{letol}\cite{let} joining the two sources.  If
one eliminates the TLS we also kill the CTCs and CTGs. One can argue
that physically this is not armful because one can thought this
singularities as the mathematical representation of a mechanical
device to keep the two particles fixed. But, since on a TLS the
geometry of the spacetime is no longer Riemannian these singularities
are rather pathological.
 
In this comment we show that we can have CTCs in a spacetime without
 TLS when the point particles (the source of the spacetime) are not equal.
This makes the appearance of CTCs in the Bonnor-Ward spacetime more
worrisome.

The Bonnor-Ward spacetime is a particular case of the Perj\'es-Israel-Wilson
(PIW) metric \cite{perjes} \cite{iw},
\begin{equation}
ds^2=-f^{-1}h_{mn}dx^m dx^n+f(\omega_m dx^m+dt)^2,
\label{metric}
\end{equation}
where the three dimensional positive definite tensor $h_{mn}$ has zero
Ricci tensor and it is usually taken as the  three dimensional
Euclidean metric in cylindrical coordinates.
The punctual sources of the  PIW spacetime are named
Perjeons \cite{bonnorwd}. A Perjeon has mass
$m$, electric charge $e$, classical angular momentum $\vec{h}$ and
magnetic moment $\vec{\mu}$ related, in relativistic units,
 by the equations: $m=\epsilon\, e$ and $\vec{h}=\epsilon\,
 \vec{\mu}$, where $\epsilon=\pm1$.

We shall consider the particular case of the Bonnor-Ward solution described in
\cite{bonnorst} with a small modification. In the definition of
function $\Omega$ ($\omega_a= -\rho^2\Omega\,\delta_a^{\varphi}$) we 
impose that $\alpha \equiv m_1\mu_2+m_2\mu_1=0$.
In this case, the Bonnor-Ward solution does not present TLS.
Note that in the cases of CTCs and CTG studied in \cite{bonnorwd}
 $\alpha\not=0$.

A closed curve $\gamma$ in this spacetime is given in its parametric
form by, $ t = t_0$,\;$ \rho = \rho_0$, \;$\varphi \in [0,2\pi]$,\;
$z=z_0$, where $t_0$, $\rho_0$ and $z_0$ are constants. This closed
curve is timelike when $g_{\varphi\varphi} > 0$, i.e., when
$\Omega^2f^2 \rho^2-1>0$. The four-acceleration $a^{\mu}$ of $\gamma$
has two nonzero components, $a^{\rho}$ and $a^z$.

In \cite{bonnorst} it is shown that when $m_1=m_2$, $\mu_1=-\mu_2$ and
$z_0=0$, we have $\alpha=0$, but in this case we have
$g_{\varphi\varphi}(\rho_0,z_0)=-\rho_0^2/f$.  Therefore the closed curve
$\gamma$ is not timelike. If we weaken these restrictions, for instance,
 taking $m_2=n\,m_1$, $\mu_2=-n\,\mu_1$ with $n\ne 1$, we have $\alpha=0$ 
and it is possible to find CTCs. 

The linear stability of a curve $\gamma$ is studied by
adding a perturbation ${\bf \xi}$ to $ (t_0 ,\rho_0, \varphi, z_0)$ 
and analyzing the behavior of
this perturbation through the modified equation of geodesic deviation
presented in ~\cite{rosalet}. For the CTC $\gamma$ we have that this
equation reduces to a  system of  second order ordinary differential equation 
with constant coefficients. This system has bound solutions  (periodic
modes) when their eigenvalues, $\lambda$, are pure imaginary numbers, 
($\lambda_1=\pm\,\imath\,\kappa_1,\;  \lambda_2=\pm\,\imath\,\kappa_2$). The explicit forms
of $\kappa_1$ and $\kappa_2$, depending on the distance $a$, the parameters
$m_1, \mu_1, \; n $, and  the position $(\rho_0,z_0)$, are rather cumbersome and will be presented elsewhere.

We find that when the parameters are: $a=5$, $\rho_0=12$, $m_1=0.1$,
$n=0.01$, $\mu_1=1800$, and $z_0=-0.1$, the closed curve $\gamma$ is
timelike ($g_{\varphi\varphi}=254.03612$) and stable
($\kappa_1=0.1712868682$ and $\kappa_2=0.9902631725$). For this
example the values of acceleration are: $a^{\rho}=0.1623818860$, and
$a^z=0.1737327461$. We get another example of a stable CTC when
$a=6$, $\rho_0=8$, $m_1=0.1$, $n=0.01$, $\mu_1=209.32$, and $z_0=0$. For
these values of the parameters, we have
$g_{\varphi\varphi}=102.0184576$, $\kappa_1=0.09159526892$, 
$\kappa_2=0.4104757936$, $a^{\rho}=0.1718928982$ and
$a^z=0.1719035561$. A third example of stable CTC  is obtained when $a=3$, $\rho_0=9.39$, $m_1=0.1$,
$n=0.001$, $\mu_1=1900$, and $z_0=0$. In this case, we get $g_{\varphi\varphi}=160.9211303$,
$\kappa_1=0.5441422796$,  $\kappa_2=1.176146266$
$a^{\rho}=0.2139746920$ and $a^z=0.3157680388$. 

To find a  CTC $\gamma$ in this spacetime is not difficult, we only need 
$g_{\varphi\varphi}>0$. But, to get a stable one ($\lambda^{2}_1<0$, and 
$\lambda^{2}_2<0$) is not so simple. We need to have one positive function
and two negative ones, our  numerical experiments indicates that these  
seldom occurs.

 The examples of CTC with no TLS presented in this comment
indicates a possible existence   of closed timelike geodesics 
in the Bonnor-Ward solution. To find a explicit example of  these
curves is not an easy task,  we need: $g_{\varphi\varphi}>0$, and $a^{\rho}= a^z=0$. If we add the condition of stability:  $\lambda^{2}_1<0$, and 
$\lambda^{2}_2<0$ (pure imaginary eivenvalues), we believe that will be almost impossible to satisfy all these conditions.
In summary, we conjecture that  stable closed timelike geodesics in the Bonnor-Ward solution without TLS do no exist.

Finally, we want to repport that we looked for closed timelike geodesics in a perturbed Bonnor-Ward solution. We found that our multipolar perturbations 
introduces TLS.

\vspace{5mm}

V.M.R. thanks the hospitality of DMA-IMECC-UNICAMP. P.S.L. thanks the
partial financial support of FAPESP and CNPq.


\end{document}